\documentclass[aps, prd, reprint, nofootinbib, longbibliography]{revtex4-1}
\usepackage{graphics,graphicx,epsfig}
\usepackage{amsmath, amssymb, color}
\usepackage[subnum]{cases}
\usepackage{setspace}

\begin{document}

\title{Spectra of conditionalization and typicality in the multiverse}
\author{Feraz Azhar}
\email[Email address: ]{feraz.azhar@alumni.physics.ucsb.edu}
\affiliation{Department of History and Philosophy of Science, University of Cambridge, Free School Lane, Cambridge, CB2 3RH, United Kingdom}

\date{\today}

\begin{abstract}
An approach to testing theories describing a multiverse, that has gained interest of late, involves comparing theory-generated probability distributions over observables with their experimentally measured values. It is likely that such distributions, were we indeed able to calculate them unambiguously, will assign low probabilities to any such experimental measurements. An alternative to thereby rejecting these theories, is to conditionalize the distributions involved by restricting attention to domains of the multiverse in which we might arise. In order to elicit a crisp prediction, however, one needs to make a further assumption about how typical we are of the chosen domains. In this paper, we investigate interactions between the spectra of available assumptions regarding both conditionalization and typicality, and draw out the effects of these interactions in a concrete setting; namely, on predictions of the total number of species that contribute significantly to dark matter. In particular, for each conditionalization scheme studied, we analyze how correlations between densities of different dark matter species affect the prediction, and explicate the effects of assumptions regarding typicality. We find that the effects of correlations can depend on the conditionalization scheme, and that in each case atypicality can significantly change the prediction. In doing so, we demonstrate the existence of overlaps in the predictions of different ``frameworks" consisting of conjunctions of theory, conditionalization scheme and typicality assumption. This conclusion highlights the acute challenges involved in using such tests to identify a preferred framework that aims to describe our observational situation in a multiverse.
\end{abstract}

\maketitle

\section{Introduction}\label{SEC:Introduction}

A central concern regarding contemporary cosmological theories that describe a multiverse, such as those involving inflationary scenarios~\cite{steinhardt_83, vilenkin_83, linde_83, linde_86a, linde_86b}, possibly in combination with the string theory landscape~\cite{bousso+polchinski_00, kachru+al_03, freivogel+al_06, susskind_07}, is: how do we elicit testable predictions from these theories? A particularly natural class of predictions are those derived from theory-generated probability distributions over observables, such as parameters of the standard models of particle physics and cosmology, or indeed observables generated from the outcomes of experiments we have yet to perform. But the task of extracting such predictions has been elusive. 

It is expected, owing to the variety of conditions that are likely to obtain in any multiverse scenario, that a theory-generated probability of our observations in such a scenario will turn out to be low. In this case, short of disfavoring (all) such theories, one can restrict attention to domains in the multiverse in which our observational situation might obtain, and then compare the new (renormalized) probability distribution with our observations.

As described by~\citet{aguirre+tegmark_05} (see also~\cite{aguirre_07}), this process of \emph{conditionalization} can occur in a variety of different ways. One possibility, termed the ``bottom-up" approach, is to not conditionalize one's distribution at all, and corresponds to accepting the raw theory-generated probability distribution as the primary means of generating a prediction. At the opposite end of the spectrum, ``top-down" conditionalization restricts attention to domains that share all observational features that we have thus far measured, except for the quantity whose value we are aiming to predict~\cite{weinstein_06, garriga+vilenkin_08, srednicki+hartle_10, hartle+hertog_13, hartle+hertog_15}. Intermediate approaches between these two ends of the spectrum propose to conditionalize on some characterization of our observational situation without demanding that all (relevant) known features be included. This last approach is termed ``anthropic", and can be thought of as according with Carter's ``weak anthropic principle"~\cite{carter_74} (see~\citet{hartle_07} for a clear discussion).

There are, as one might expect, inherent difficulties in implementing either anthropic or top-down conditionalization schemes, arising from how best to characterize ``us" in the anthropic case, or how to characterize our observational situation in a practicable way in the top-down case. Even if one is able to address these issues, there remains a further assumption that needs to be made in order to extract a crisp prediction. This amounts to an assumption regarding how typical we are of the domains that these conditionalization schemes explicitly restrict attention to. For a renormalized probability distribution function (and, indeed, for distributions exhibiting the appropriate shape), this amounts to an assumption about how far away from the peak of a distribution we can allow our observations to be, while still taking those observations to have been predicted by the conjunction of theory, conditionalization scheme and typicality assumption---a conjunction we will refer to as a \emph{framework}, in accord with the terminology of~\citet{srednicki+hartle_10}.

What typicality assumptions one should support is controversial. There are essentially two camps: those who assert that we should always assume typicality in the context of an appropriately conditionalized theory, that is, those who support the ``principle of mediocrity"~\cite{gott_93, vilenkin_95, page_96, bostrom_02, garriga+vilenkin_08}, and those who see typicality as an assumption that can be subject to error, and that therefore we should not necessarily demand typicality, whatever our specification of the conditionalized theory~\cite{hartle+srednicki_07, smolin_07, srednicki+hartle_10, azhar_14, azhar_15}. If one allows for the latter possibility, then one is faced with a spectrum of possible typicality assumptions. Given some theory, it is then by some appropriate choice in \emph{each} of the spectra of conditionalization \emph{and} typicality that one must extract predictions. 

In this paper, we take seriously the need to consider the existence of these two spectra, and investigate the manner in which they interact, for a range of conditionalization schemes and typicality assumptions. This investigation is carried out in the concrete context of an attempt to predict the total number of species that contribute significantly to dark matter. In particular, we extend the work of~\citet{aguirre+tegmark_05} by considering cases where probability distributions over densities of dark matter species can be correlated, and then analyze the effects of bottom-up (Sec.~\ref{SEC:BU}), top-down (Sec.~\ref{SEC:TD}), and anthropic (Sec.~\ref{SEC:AR}) conditionalization schemes, in addition to the effects of atypicality in each of these cases. We find that (i) atypicality can significantly change the prediction in each case studied in such a way that (ii) different frameworks can overlap, as regards their predictions; that is, different frameworks can lead to the same prediction for the number of dominant species of dark matter (Sec.~\ref{SEC:Overlaps}). These results leave open the challenge, in more realistic settings, of constructing these sets of equivalent frameworks (as judged by the equivalence of their predictions); while also highlighting how difficult it may be to use such tests to identify a preferred framework. 

\subsection{The general cosmological setting}\label{SEC:GenCosmo}

We begin by briefly outlining the general cosmological scenario within which we will be working. The current favored theory regarding the composition of dark matter does not rule out the possibility of multiple (new, non-baryonic) particle species contributing to the total dark matter density~\cite{aguirre+tegmark_05, bertone+al_05, tegmark+al_06}. The general argument of this paper will build upon this possibility: we will assume that some theory $\mathcal{T}$ describes a multiverse consisting of distinct domains, in each of which a total of $N$ distinct species of dark matter can exist, but where the relative contributions of each of these species to the total dark matter density can vary from one domain to the next. The densities of each of these components will be given by a dimensionless dark-matter-to-baryon ratio (we will also assume that the density of baryons can vary from one domain to the next), with the density of component $i$ given by $\eta_{i}\equiv\Omega_{i}/\Omega_{\textrm{b}}$, so that the densities of all $N$ components are represented by $\vec{\eta} = (\eta_{1}, \eta_{2}, \dots, \eta_{N})$. Note that our observations currently constrain the \emph{total} dark matter density $\eta_{\textrm{obs}}\equiv\sum_{i=1}^{N}\eta_{i}$. From results recently released by the \emph{Planck} collaboration, this quantity can be shown to be $\eta_{\textrm{obs}}\approx 5$~\cite{planck_15_CP}. 

The space in which $\vec{\eta}$ will take values will be referred to as ``parameter space". The variation of this vector of densities from one domain to the next is described by a probability distribution $P(\vec{\eta}\,|\mathcal{T})$. The construction of such probability distributions is a difficult, open problem, and to make progress we will specify simple, example distributions as we proceed. We begin then by considering the least restricted case: that of bottom-up conditionalization.

\section{Bottom-up conditionalization}\label{SEC:BU}

According to bottom-up conditionalization, one assumes that the raw probability distribution $P(\vec{\eta}\,|\mathcal{T})$ constitutes the primary means of generating predictions. We will assume, following the general line of argument in~\citet{aguirre+tegmark_05}, that in principle, the range in which each of the  component densities $\eta_{i}$ in $\vec{\eta}$ could take values is large (and is the same for each species $i$). Assume also that the joint probability distribution $P(\vec{\eta}\,|\mathcal{T})$ is unimodal, that is, has a single peak that could fall anywhere in the range over which $P(\vec{\eta}\,|\mathcal{T})$ could possibly be significant [in this section, we make no assumptions regarding the nature of any correlations between the component densities for any particular $P(\vec{\eta}\,|\mathcal{T})$]. In the absence of any further information, we are interested in the following two questions: (i) how many of the $N$ components share the highest occurring density, and (ii) how does this prediction depend upon the assumption of typicality?

To make this problem tractable, let us discretize the range over which each of the densities could take values into $M$ equal-sized bins, such that the central density of each bin is significantly different from its neighbors. We thus have an $N$-dimensional grid containing $M^{N}$ boxes, where we assume the peak of $P(\vec{\eta}\,|\mathcal{T})$ is equally likely to fall into any box, and we are interested first [i.e., in (i) above], in the probability that a total of $j$ of the $N$ components share the highest occupied density box. To be clear, a particular $\mathcal{T}$ will indeed give rise to a single $P(\vec{\eta}\,|\mathcal{T})$ where the peak of this distribution will have a single location in the grid---we are looking into the situation where we have no further information about the location of this peak, and are interested in predictions about where the peak will lie, assuming that it is equally likely to fall into any of the boxes we have constructed. 

It should be intuitively clear that for $N\geq 2$ and high enough $M$ (i.e., $M\gg N$), the chance of the peak falling along the equal density diagonal of the $N$-dimensional grid is small. We can formalize this intuition with the following amendment to the corresponding argument in~\citet[Sec.~3.2]{aguirre+tegmark_05}. In this amended argument, the final result we obtain for the probability of $j$ components sharing the highest occupied density box, namely $\mathcal{P}(j)$, is different from their result [their equation (1)], but the overall conclusion of the analysis of bottom-up conditionalization remains the same. 

The problem as stated in the previous paragraph can be recast in the following (dimensionally-reduced) form where we consider $M$ distinguishable bins, corresponding to the discretized densities in the range over which any dark matter species can take values, and $N$ distinguishable balls, where the $i$'th ball represents the peak of the $i$'th marginal distribution $P_{i}(\eta_{i}\,|\mathcal{T})$. Our assumption that the peak of $P(\vec{\eta}\,|\mathcal{T})$ is equally likely to fall into any box is equivalent to the statement that the probability of any ball falling into any bin is the same. Let $\mathcal{P}(j)$ represent the probability that exactly $j$ of the $N$ balls fall into the highest occupied density bin. Then the following closed-form expression, obtained through a simple counting argument gives us the required probability $\mathcal{P}(j)$:
\begin{equation}\label{EQN:Probj}
\mathcal{P}(j) = \frac{1}{M^{N}}\binom{N}{j}\left[\sum_{k=1}^{M-1}k^{N-j}+\delta_{j,N}\right],
\end{equation}
where $\delta$ is the Kronecker delta function. 

To understand where this result comes from, consider the case where some $j<N$ balls share the highest occupied density bin. If that bin is the $k$'th from the lowest density bin of the $M$ possible bins (where $1\leq k \leq M-1$), then all the remaining $N-j$ balls can be arranged in the $k$ lower density bins in $k^{N-j}$ ways. The sum in Eq.~(\ref{EQN:Probj}) corresponds to the sum over all possible choices of $k$. The prefactor $\binom{N}{j}$ just counts the number of ways of selecting exactly $j$ of the $N$ balls. This product is then divided by the total number of possible arrangements of balls in bins, i.e., $M^{N}$, giving the appropriate probability. The Kronecker delta function keeps track of the particular case where $j=N$, in which case there exists an extra arrangement in which the $j$ balls sharing the highest occupied density bin (i.e., all $N$ of them) can indeed be placed in the lowest density bin. It is straightforward to show that this distribution is appropriately normalized: $\sum_{j=1}^{N}\mathcal{P}(j) = 1$.

Our intuition that the chance is small of the peak of $P(\vec{\eta}\,|\mathcal{T})$ falling along the equal density diagonal in the $N$-dimensional grid (i.e., in the dimensionally-reduced description in the paragraph above, of all $N$ balls falling into the same bin), suggests that $\langle j \rangle\sim 1$. We will outline how for the most likely relative values of $N$ and $M$, this is indeed the case. We find for $\langle j \rangle$:
\begin{eqnarray}\label{EQN:AvgJ}
\langle j \rangle &\equiv& \sum_{j=1}^{N}j\mathcal{P}(j) \nonumber \\
				    &=& \frac{1}{M^{N}}\sum_{j=1}^{N}j\binom{N}{j}\left[\sum_{k=1}^{M-1}k^{N-j}+\delta_{j,N}\right]\nonumber \\
				    &=& \frac{1}{M^{N}}\sum_{k=1}^{M-1}\sum_{j=1}^{N}j\binom{N}{j}k^{N-j}+\frac{N}{M^{N}}\nonumber \\
				    &=& \frac{N}{M^{N}}\sum_{k=1}^{M-1}\sum_{j=1}^{N}\binom{N-1}{j-1}k^{N-j}+\frac{N}{M^{N}}\nonumber \\
				    &=& \frac{N}{M^{N}}\left[\sum_{k=1}^{M-1}(1+k)^{N-1}+1\right]\nonumber \\
				    &=& \frac{N}{M^{N}}\sum_{k=1}^{M}k^{N-1}, 
\end{eqnarray}
where the binomial theorem has been used in obtaining the fifth line. For $M\gg N$, one can show that $\langle j \rangle \sim 1$ (formally: for fixed $N$, $\lim_{M \to \infty}\langle j \rangle = 1$). In the case where $N\gg M$, $\langle j \rangle$ can take values much greater than 1, that is, it is possible for multiple components to dominate (formally: for fixed $M$, $\lim_{N \to \infty}\langle j \rangle = \infty$). The upshot is that as long as the range over which each of the dark matter densities can take values, namely $M$, is much larger than the total number of dark matter species under consideration, namely $N$, the average number of species sharing the highest occurring density will be 1. 

This result has been derived under the assumption of typicality, in that the peak of the joint distribution dictates the prediction. To be clear, the average in Eq.~(\ref{EQN:AvgJ}) is taken over all possible locations of the peak of $P(\vec{\eta}\,|\mathcal{T})$, and assumes that we do not in fact know where this might be. Of course, for any fixed $\mathcal{T}$, the peak of the joint distribution $P(\vec{\eta}\,|\mathcal{T})$ will be located in a single box, and the argument following Eq.~(\ref{EQN:AvgJ}) indicates that this box will probably correspond to a  single dominant dark matter component. 

Indeed, irrespective of the location of the peak of any particular distribution $P(\vec{\eta}\,|\mathcal{T})$, there will presumably exist directions in parameter space in which more than a single component would contribute significantly. Therefore atypicality can lead to a range of different predictions for the total number of species that contribute significantly. Of course, we are constrained here in that only those theories (and their associated typicality assumptions) that predict a total density that agrees with our observed value, i.e., that satisfy $\sum_{i=1}^{N}\eta_{i}=\eta_{\textrm{obs}}\approx 5$, would be favored. Nevertheless, the combination of bottom-up conditionalization and atypicality presents us with a large amount of freedom regarding predictions that may arise.

\section{Top-down conditionalization}\label{SEC:TD}

Our arguments thus far have been rather general, and it will be instructive in what follows to restrict attention to particular distributions so as to extract more concrete predictions. We turn now to the most restrictive type of conditionalization scheme---that of top-down conditionalization.

Consider, again, the case where we have a total of $N$ species of dark matter and we are interested in ascertaining the total number of species that contribute significantly to the total observed dark matter density $\eta_{\textrm{obs}}$. Assume that the joint probability distribution function given some theory $\mathcal{T}$ is locally Gaussian (near $\eta_{\textrm{obs}}$), with
\begin{equation}\label{EQN:CorrGauss}
P(\vec{\eta}\,|\mathcal{T}) \propto \exp\left\{-\frac{1}{2}\sum_{i,j = 1}^{N}(\eta_{i}-\eta_{i}^{\star})(\mathcal{C}^{-1})_{ij}(\eta_{j}-\eta_{j}^{\star})\right\},
\end{equation}
where each component has substantial probability near $\eta_{\textrm{obs}}$, i.e., for each $i$, $\eta_{i}^{\star}\sim\eta_{\textrm{obs}}$ (where we allow for some tolerance in the precise relationship between $\eta_{i}^{\star}$ and $\eta_{\textrm{obs}}$ here), and $\mathcal{C}$ is the covariance matrix (a symmetric, positive definite, $N\times N$ matrix). This matrix, of course, encodes potential correlations between each of the components. Note that if $\mathcal{C}_{ij}\propto\delta_{ij}$ then the right hand side of Eq.~(\ref{EQN:CorrGauss}) reduces to a product of independent Gaussians.

Top-down conditionalization in this scenario amounts to demanding that the prediction extracted from this distribution agrees with the totality of our data regarding dark matter (see Sec.~\ref{SEC:GenCosmo}); namely, that the sum over the densities of dark matter components agrees with the total observed dark matter density, that is:
\begin{equation}\label{EQN:TDConstraint}
\sum_{i=1}^{N}\eta_{i} = \eta_{\textrm{obs}}.
\end{equation}
Generating a prediction under the assumption of typicality amounts to finding $\vec{\eta}$ such that Eq.~(\ref{EQN:CorrGauss}) is maximized, subject to Eq.~(\ref{EQN:TDConstraint}), and it is to this task that we now turn. 

\subsection{Typicality}\label{SEC:TDGaussTyp}

We proceed as in~\citet{aguirre+tegmark_05}, and focus on the constrained optimization problem in which we optimize the logarithm of the distribution $P(\vec{\eta}\,|\mathcal{T})$, that is, we aim to maximize
\begin{equation}
I(\vec{\eta}) \equiv \ln P(\vec{\eta}\,|\mathcal{T}) - \lambda\sum_{i=1}^{N}\eta_{i}, 
\end{equation}
where $\lambda$ is a Lagrange multiplier. A quick calculation shows that for each $k$, $\partial I(\vec{\eta})/\partial\eta_{k}=0$ when 
\begin{equation}
\sum_{i = 1}^{N}(\eta_{i}-\eta_{i}^{\star})(\mathcal{C}^{-1})_{ik}=-\lambda.
\end{equation}
Multiplying through by $\mathcal{C}_{kj}$ and summing over $k$ gives
\begin{equation}\label{EQN:IntLambda}
\eta_{j}-\eta_{j}^{\star} = -\lambda\sum_{k=1}^{N}\mathcal{C}_{kj}. 
\end{equation}
By summing the $N$ equations implicit in Eq.~(\ref{EQN:IntLambda}), rearranging, and using the constraint [Eq.~(\ref{EQN:TDConstraint})], we can solve for the Lagrange multiplier:
\begin{equation}\label{EQN:FullLambda}
\lambda = \frac{\sum_{j=1}^{N}\eta_{j}^{\star}-\eta_{\textrm{obs}}}{\sum_{k,j = 1}^{N}\mathcal{C}_{kj}}.
\end{equation}
Substituting Eq.~(\ref{EQN:FullLambda}) into Eq.~(\ref{EQN:IntLambda}) we find that the maximum of $P(\vec{\eta}\,|\mathcal{T})$ subject to the top-down constraint occurs at 
\begin{equation}\label{EQN:GeneralEtaMax}
\eta_{i}=\eta_{i}^{\star}+\left(\frac{\eta_{\textrm{obs}}-\sum_{j=1}^{N}\eta_{j}^{\star}}{\sum_{k,l=1}^{N}\mathcal{C}_{kl}}\right)\sum_{m=1}^{N}\mathcal{C}_{mi}.
\end{equation}
A judicious choice of the $\eta_{i}^{\star}$'s and/or the sums of the columns of the covariance matrix $\mathcal{C}$, therefore, can lead to substantial contributions by less than all $N$ species. 

However, this conclusion is overturned, i.e., all $N$ species contribute equally, in the case where $(i)$ no symmetries are broken with regard to the location of the peak of the joint probability distribution, namely, if for each $i$,
\begin{equation}\label{EQN:EtaUniform}
\eta_{i}^{\star} = \bar{\eta}
\end{equation}
for some $\bar{\eta}\sim\eta_{\textrm{obs}}$, and $(ii)$ we choose an appropriate functional form for the covariance matrix. In particular, let us assume that the $N\times N$ covariance matrix is given by $\tilde{\mathcal{C}}$ where
\begin{equation}\label{EQN:CovarianceMatrix}
\tilde{\mathcal{C}} = \bar{\sigma}^{2}
 \begin{pmatrix}
  1 & \alpha & \cdots & \alpha \\
  \alpha & 1 & \cdots & \alpha \\
  \vdots  & \vdots  & \ddots & \vdots  \\
  \alpha & \alpha & \cdots & 1 
 \end{pmatrix}
\end{equation}
with $\alpha\in(-(N-1)^{-1}, 1)$, so that $\tilde{\mathcal{C}}$ is indeed positive definite, and $\bar{\sigma}$ is a free parameter. This choice fixes all variances to be the same, and all pairs of covariances to be the same, that is,
\begin{eqnarray}
\langle(\eta_{i}-\bar{\eta})^2 \rangle &=& \bar{\sigma}^{2}\;\;\;\forall i,\label{EQN:Var} \\
\langle(\eta_{i}-\bar{\eta})(\eta_{j}-\bar{\eta}) \rangle &=& \alpha\,\bar{\sigma}^{2}\;\;\;\forall i\neq j.\label{EQN:Covar}
\end{eqnarray}
Then substituting Eq.~(\ref{EQN:EtaUniform}) and Eq.~(\ref{EQN:CovarianceMatrix}) into Eq.~(\ref{EQN:GeneralEtaMax}) gives, for each $i$,
\begin{eqnarray}
\eta_{i} &=& \bar{\eta}+\left[\frac{\eta_{\textrm{obs}}-N\bar{\eta}}{N\bar{\sigma}^{2}+N(N-1)\alpha\bar{\sigma}^{2}}\right]\left[\bar{\sigma}^{2}+(N-1)\alpha\bar{\sigma}^{2}\right]\nonumber\\
&=&\bar{\eta}+\frac{1}{N}\left(\eta_{\textrm{obs}}-N\bar{\eta}\right)\nonumber\\
&=& \frac{1}{N}\eta_{\textrm{obs}}. \label{EQN:EqualContribution}
\end{eqnarray}
So for a rather simple probability distribution [defined by Eqs.~(\ref{EQN:CorrGauss}),~(\ref{EQN:EtaUniform}) and~(\ref{EQN:CovarianceMatrix})] with (marginal) probability distributions over distinct dark matter species that are, in effect, the same: the most probable, i.e., most \emph{typical}, solution to the ensuing constrained maximization problem is that all $N$ components contribute equally to the total dark matter density. 

This extends the argument of~\citet[Sec.~3.3]{aguirre+tegmark_05} to the case where correlations are now explicitly built into the joint probability distribution $P(\vec{\eta}|\mathcal{T})$. Note that the results above include the case of probabilistically independent dark matter species, that is, for the special case of $\alpha = 0$ we would again find that  Eq.~(\ref{EQN:EqualContribution}) holds. 

So how does atypicality affect the picture? We will now show that under an assumption of atypicality, there exists a region in parameter space where just a single dark matter component dominates. That is, atypicality can dramatically change the prediction. 

\subsection{Atypicality}\label{SEC:TDGaussAtyp}

We focus on the case where the underlying probability distribution is given by Eq.~(\ref{EQN:CorrGauss}), but is subject to the assumption that the peak of the distribution does not privilege any dark matter species, that is, Eq.~(\ref{EQN:EtaUniform}) holds, and the covariance matrix is given by Eq.~(\ref{EQN:CovarianceMatrix})---these latter two assumptions are made in order to mitigate obvious biases in the search for a single dominant species. We will label the resulting distribution $P_{\textrm{E}}(\vec{\eta}\,|\mathcal{T})$ (`E' for `equal'); so that, from Eqs.~(\ref{EQN:CorrGauss}),~(\ref{EQN:EtaUniform}) and~(\ref{EQN:CovarianceMatrix}),
\begin{equation}\label{EQN:CorrGaussEtaUniform}
P_{\textrm{E}}(\vec{\eta}\,|\mathcal{T}) \propto \exp\left\{-\frac{1}{2}\sum_{i,j = 1}^{N}(\eta_{i}-\bar{\eta})(\tilde{\mathcal{C}}^{-1})_{ij}(\eta_{j}-\bar{\eta})\right\}. 
\end{equation}
To highlight the effects of atypicality, we will look for a region in parameter space away from the maximum of the probability on the constraint surface [which occurs when Eq.~(\ref{EQN:EqualContribution}) is satisfied], while remaining on the constraint surface [i.e., respecting Eq.~(\ref{EQN:TDConstraint})]. So, from Eq.~(\ref{EQN:EqualContribution}), the maximum of the probability on the constraint surface, denoted by $P^{\textrm{MAX}}$,  is given by
\begin{equation}
P^{\textrm{MAX}} \equiv P_{\textrm{E}}\left(\left\{{\eta}_{i} = \frac{1}{N}\eta_{\textrm{obs}}\right\}_{i=1}^{N}\middle|\mathcal{T}\right). 
\end{equation}
Our excursion on the constraint surface will explore the possibility of just a single species dominating, and we will take that species to be the first species (though of course, nothing physical depends on this choice). Hence, following~\citet[Sec. 3.2.2]{azhar_14}, we remain on the constraint surface by demanding that 
\begin{eqnarray}
\eta_{1}^{\prime} &=& \epsilon\frac{1}{N}\eta_{\textrm{obs}},\label{EQN:eta1Param}\\
\eta_{j}^{\prime} &=& \left(1-\frac{\epsilon}{N}\right)\frac{1}{N-1}\eta_{\textrm{obs}}\;\;\;\;\;\forall j\neq 1\label{EQN:etajParam}, 
\end{eqnarray}
for $0\leq\epsilon\leq N$. Notice that $\sum_{i=1}^{N}\eta_{i}^{\prime}=\eta_{\textrm{obs}}$, and species $1$ dominates when $\epsilon\to N$. The parameterization is chosen in such a way that any excess in 
$\eta_{1}^{\prime}$ over the most probable value, $\eta_{1}^{\prime} =\frac{1}{N}\eta_{\textrm{obs}}$, is drawn equally from among the remaining $N-1$ components. We gauge the degree of typicality (and thereby the degree of atypicality) by the ratio of the probability $P_{\textrm{E}}(\vec{\eta}\,^{\prime}\,|\mathcal{T})$, with $\vec{\eta}\,^{\prime}\equiv\vec{\eta}\,^{\prime}(\epsilon)$ using the parameterization in Eqs.~(\ref{EQN:eta1Param}) and~(\ref{EQN:etajParam}), to $P^{\textrm{MAX}}$. A lengthy (but straightforward) calculation reveals that this ratio is given by
\begin{equation}\label{EQN:AtypRatio}
\frac{P_{\textrm{E}}(\vec{\eta}\,^{\prime}(\epsilon)\,|\mathcal{T})}{P^{\textrm{MAX}}} = \exp\left\{-\frac{{\eta_{\textrm{obs}}}^{2}}{2\bar{\sigma}^2(1-\alpha)}\frac{(\epsilon-1)^2}{N(N-1)}\right\}, 
\end{equation}
where as before, $\alpha$ describes correlations between different species [cf. Eq.~(\ref{EQN:Covar})]. We point out two interesting features of the result in Eq.~(\ref{EQN:AtypRatio}):
\begin{itemize}
\item[(i)] The dominance of species 1 indeed relies on an assumption of atypicality. To illustrate this, consider the case where there exist two total dark matter components ($N=2$) where the density of species 1 is three times that of species 2 ($\epsilon = 3 N/4$). If we further assume that $\bar{\sigma} = \eta_{\textrm{obs}}/5$ and $\alpha = 1/2$, the degree of typicality that achieves this dominance is small: ${P_{\textrm{E}}(\vec{\eta}\,^{\prime}(\epsilon)\,|\mathcal{T})/P^{\textrm{MAX}}}=\exp(-25/8)\approx 0.04$;
\item[(ii)] Correlations affect the degree of typicality required to achieve the same dominance. Hence in the example in $\textrm{(i)}$ above: setting $\alpha = 0$ [i.e., no correlations, cf. Eq.~(\ref{EQN:Covar})], while keeping all other parameters the same, gives ${P_{\textrm{E}}(\vec{\eta}\,^{\prime}(\epsilon)\,|\mathcal{T})/P^{\textrm{MAX}}}=\exp(-25/16)\approx 0.21$. 
\end{itemize}
In this way, an assumption of atypicality can change the prediction from $N$ equally dominant components to a single dominant component. 

\subsection{Non-Gaussianities and typicality assumptions}

Thus far, we have focussed on the Gaussian case. But it is interesting to explore the situation where we explicitly break this Gaussianity---for as we will show, this can change predictions under various assumptions regarding typicality. In particular, we will show that our earlier conclusion, which established equal contributions to the total dark matter density under typicality, as expressed in Eq.~(\ref{EQN:EqualContribution}), can be overturned when we break Gaussianity. 

We will establish this result numerically and by construction, in the case where $N=2$. We will assume that the underlying distribution $Q(\eta_{1}, \eta_{2}|\mathcal{T})$ is non-Gaussian, where the non-Gaussianity is controlled by a single positive parameter $\mu$, such that $\mu=0$ recovers the Gaussian case. Assume, then, that 
\begin{widetext}
\begin{equation}\label{EQN:NonGaussDist}
Q(\eta_{1}, \eta_{2}|\mathcal{T}) \propto \left[1+\mu\sum_{k=1}^{2}(\eta_{k}-\eta_{k}^{\star})^{4}\right]\exp\left\{-\frac{1}{2}\sum_{i,j = 1}^{2}(\eta_{i}-\eta_{i}^{\star})(\tilde{\mathcal{C}}^{-1})_{ij}(\eta_{j}-\eta_{j}^{\star})\right\},
\end{equation}
\end{widetext}
where $\tilde{\mathcal{C}}$ is the two-dimensional version of Eq.~(\ref{EQN:CovarianceMatrix}), and so $\alpha\in (-1,1)$. We want to maximize $Q(\eta_{1}, \eta_{2}|\mathcal{T})$ subject to the constraint that the sum of the densities is the observed dark matter density:
\begin{equation}\label{EQN:NonGaussConstraint}
\eta_{1}+\eta_{2}=\eta_{\textrm{obs}}.
\end{equation}
So proceeding as in section~\ref{SEC:TDGaussTyp}, we wish to maximize
\begin{equation}
J(\eta_{1}, \eta_{2})\equiv \ln Q(\eta_{1}, \eta_{2}|\mathcal{T}) - \lambda\sum_{i=1}^{2}\eta_{i}
\end{equation}
where $\lambda$ is a Lagrange multiplier. Setting $\partial J(\eta_{1}, \eta_{2})/\partial\eta_{k}=0$, gives the following two equations (for $k=1,2$):
\begin{equation}\label{EQN:NonGaussSolns}
\frac{4\mu(\eta_{k}-\eta_{k}^{\star})^3}{1+\mu\sum_{i=1}^{2}(\eta_{i}-\eta_{i}^{\star})^{4}}-\sum_{j=1}^{2}(\eta_{j}-\eta_{j}^{\star})(\tilde{\mathcal{C}}^{-1})_{jk}= \lambda.
\end{equation}
The set of equations given by Eq.~(\ref{EQN:NonGaussSolns}) and Eq.~(\ref{EQN:NonGaussConstraint}) constitute three equations for the three unknowns $\{\eta_{1}, \eta_{2}, \lambda\}$. 

\begin{figure*}
\includegraphics[width=1\linewidth]{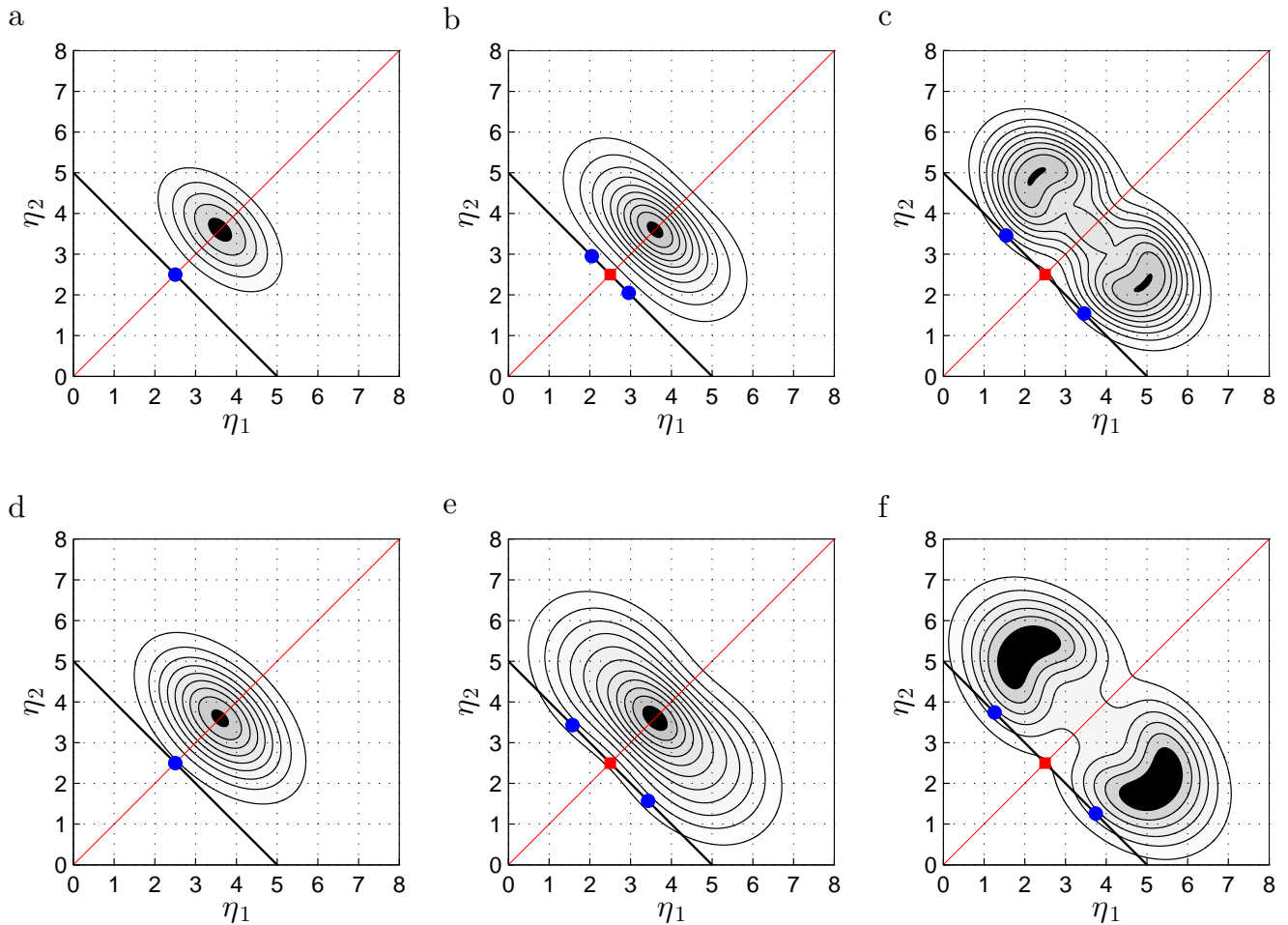}
\hfill
\caption{Contour plots of the distribution $Q(\eta_{1}, \eta_{2}|\mathcal{T})$ [see Eq.~(\ref{EQN:NonGaussDist})]. In each panel, the black line denotes the constraint surface, and the red line denotes the line of equal density. Along the constraint surface, blue circles correspond to (global) maxima and red squares to (local) minima. Note that in (b,c,e,f), pairs of maxima have the same probability in each panel. Parameters have been set as follows: $\eta_{\textrm{obs}}=5$, $\eta_{1}^{\star} = 3.6 =  \eta_{2}^{\star}$, and $\alpha = -0.5$. (a--c) $\bar{\sigma} = \eta_{\textrm{obs}}/6$ with $\mu = 0, 0.1,\textrm{and}\,1$ respectively. (d--f) $\bar{\sigma} = \eta_{\textrm{obs}}/5$ with $\mu = 0, 0.1,\textrm{and}\,1$ respectively. The Gaussian cases (a,d) exhibit a single maximum corresponding to equal contributions to the total dark matter density from the two components (as discussed in section~\ref{SEC:TDGaussTyp}). The equality of contribution is overturned in a significant way for $\mu=1$, that is (c,f), where in each case, the maxima correspond to unequal contributions and the local minimum corresponds to equal contributions.}
\label{FIG:NonGaussOptimization}
\end{figure*}

We proceed to solve these numerically, and plot the resulting solutions in Fig.~\ref{FIG:NonGaussOptimization}. 
The free parameters in the problem are $\{\mu, \eta_{1}^{\star}, \eta_{2}^{\star}, \alpha, \bar{\sigma}, \eta_{obs} \}$. Figure~\ref{FIG:NonGaussOptimization} displays results such that for each of two values of $\bar{\sigma}$ (each value corresponds to a row of the plot), the parameter controlling the non-Gaussianity, namely $\mu$, is varied over three possible values from 0 to 0.1 to 1 (from left to right). Each value corresponds to a column of the plot. We choose $\eta_{\textrm{obs}}=5$ (which is, recall, approximately the experimentally observed value), $\eta_{1}^{\star} = 3.6 =  \eta_{2}^{\star}$ (so that the distribution has weight near the experimentally observed value), and $\alpha = -0.5$ (for illustrative purposes). 

We see that in the Gaussian case ($\mu=0$, Fig.~\ref{FIG:NonGaussOptimization}a, \ref{FIG:NonGaussOptimization}d), there is a single maximum (blue circle) which occurs at $\eta_{1}= \eta_{\textrm{obs}}/2=\eta_{2}$ in accord with the general result derived earlier [Eq.~(\ref{EQN:EqualContribution})]. When we break Gaussianity we overturn this result. For smaller deviations from Gaussianity ($\mu=0.1$, Fig.~\ref{FIG:NonGaussOptimization}b,~\ref{FIG:NonGaussOptimization}e), the maxima on the constraint surface correspond to (two symmetric cases in which) one component slightly dominates over the other (with a roughly 3:2 split in Fig.~\ref{FIG:NonGaussOptimization}b and a roughly 3.4:1.6 split in Fig.~\ref{FIG:NonGaussOptimization}e). There is also a single (local) minimum on the constraint surface between these maxima, whose probability is close to theirs. This result is amplified in the case where the non-Gaussianity is stronger ($\mu=1$, Fig.~\ref{FIG:NonGaussOptimization}c,~\ref{FIG:NonGaussOptimization}f). The maxima on the constraint surface have a probability that is significantly greater than the single (local) minimum (by a factor of $\gtrsim 2$), and the dominance of one component is also greater than in the $\mu=0.1$ case, with a roughly 3.5:1.5 split in Fig.~\ref{FIG:NonGaussOptimization}c, and a roughly 3.7:1.3 split in Fig.~\ref{FIG:NonGaussOptimization}f. 

Atypicality therefore predicts either the existence of two equally contributing components [corresponding to the local minima (red squares) in Fig.~\ref{FIG:NonGaussOptimization}c,~\ref{FIG:NonGaussOptimization}f] or indeed just a single dominant component (that is more dominant than the prediction under typicality, that is, as one moves along the constraint surface towards either axis in Fig.~\ref{FIG:NonGaussOptimization}c,~\ref{FIG:NonGaussOptimization}f, say\footnote{Note that one would need to worry about boundary conditions of the distributions presented to give precise details in this case.}).

Thus we have exhibited a scenario in the top-down approach where: atypicality corresponds to equal contributions, and typicality to unequal contributions (i.e., one dominant species), to the total dark matter density. Non-Gaussianities can change the nature of the prediction. 

\section{Anthropic conditionalization}\label{SEC:AR}

Anthropic conditionalization represents an intermediate point between bottom-up and top-down approaches in that some multiverse domains are indeed excised in the computation of probabilities but the restriction is not as stringent as in the case of top-down conditionalization. 

From a calculational point of view, as discussed in~\citet{aguirre+tegmark_05}, one can implement anthropic conditionalization by adopting a weighting factor $W$  that multiplies the raw probability distribution $P(\vec{\eta}\,|\mathcal{T})$ and expresses the probability of finding domains in which we might exist, as a function of the relevant parameter we are investigating. In~\citet{aguirre+tegmark_05}, and in what follows, the assumption is made that $W\equiv W(\eta)$ is a function of the total dark matter density $\eta\equiv\sum_{i=1}^{N}\eta_{i}$, and we will look into the effects of assuming an $\eta$-dependent Gaussian fall-off for this weighting factor. That is, we will assume, following~\cite{aguirre+tegmark_05} that 
\begin{equation}\label{EQN:AnthropicFactor}
W(\eta) \propto \exp\left\{-\frac{1}{2\eta_{0}^{\,2}}\eta^2\right\}, 
\end{equation}
where we also assume that we have a way of calculating the standard deviation $\eta_{0}$.\footnote{Note that for the sake of calculational simplicity, we extend the domain of validity of the Gaussian fall-off for the anthropic weighting factor $W(\eta)$, beyond that explored in~\cite{aguirre+tegmark_05}, where this domain corresponded to $\eta > \eta_{0}$. This raises a subtlety regarding the value(s) of $\eta$ that can appropriately be considered to maximize $W(\eta)$. This is a debate that lies outside the scope of the problem considered in this paper, but would need to be addressed in a less stylized setting.} The optimization problem that implements the assumption of typicality now demands that we maximize the total probability distribution $P_{\textrm{tot}}(\vec{\eta}\,|\mathcal{T}, W)$, which takes this anthropic weighting factor into account, where
\begin{equation}\label{EQN:AnthropicTotalProb}
P_{\textrm{tot}}(\vec{\eta}\,|\mathcal{T}, W)\propto P(\vec{\eta}\,|\mathcal{T})W(\eta).
\end{equation}
We will investigate the result of doing this for the correlated Gaussian case discussed in Sec.~\ref{SEC:TD}, beginning first with the case where we do not restrict the covariance matrix. In addition, in contrast to our discussion so far, we will focus less on the equality of contribution of different components to the total dark matter density; and more on a new feature that arises exclusively in the anthropic approach: that of the determination of precisely \emph{how many} components $N$ contribute equally to the total dark matter density. 

\subsection{The optimization routine}

We assume again, that there are $N$ possible species of dark matter and that we need to find the value of $\vec{\eta}$ such that $P_{\textrm{tot}}(\vec{\eta}\,|\mathcal{T}, W)$ is maximized, where, substituting Eq.~(\ref{EQN:CorrGauss}) and Eq.~(\ref{EQN:AnthropicFactor}) into Eq.~(\ref{EQN:AnthropicTotalProb}), we have
\begin{widetext}
\begin{equation}
P_{\textrm{tot}}(\vec{\eta}\,|\mathcal{T}, W)\propto \exp\left\{-\frac{1}{2}\sum_{i,j = 1}^{N}(\eta_{i}-\eta_{i}^{\star})(\mathcal{C}^{-1})_{ij}(\eta_{j}-\eta_{j}^{\star})\right\}\exp\left\{-\frac{1}{2\eta_{0}^{\,2}}\eta^2\right\}.
\end{equation}
\end{widetext}
For each $k$, setting $\partial\ln P_{\textrm{tot}}(\vec{\eta}\,|\mathcal{T}, W)/\partial\eta_{k}=0$ gives 
\begin{equation}
\sum_{i=1}^{N}(\eta_{i}-\eta_{i}^{\star})(\mathcal{C}^{-1})_{ik} = -\frac{1}{\eta_{0}^{\,2}}\eta. 
\end{equation}
Multiplying through by $\mathcal{C}_{kj}$, summing over $k$ and rearranging, we find
\begin{equation}
\eta_{j} = \eta_{j}^{\star} -\frac{1}{\eta_{0}^{\,2}}\eta\sum_{k=1}^{N}\mathcal{C}_{kj}. 
\end{equation}
As in section~\ref{SEC:TDGaussTyp}, choosing the $\eta_{j}^{\star}$'s and/or the sums of the columns of the covariance matrix appropriately, one can find contributions to the total dark matter density such that all species do not contribute equally. 

However, under the assumption that $\eta_{j}^{\star}=\bar{\eta}$ for all $j$ [i.e., Eq.~(\ref{EQN:EtaUniform})], and that the covariance matrix is given by Eq.~(\ref{EQN:CovarianceMatrix}) say, we recover the case of equal contributions discussed above. In particular, we obtain, for each $j$,
\begin{equation}
\eta_{j} = \bar{\eta}-\frac{1}{\eta_{0}^{\,2}}\eta\left[1+(N-1)\alpha\right]\bar{\sigma}^2,
\end{equation}
where the right-hand side does not depend on $j$. The solution to these equations is just $\eta_{j} = \gamma$, say, so that $\eta\equiv\sum_{i=1}^{N}\eta_{i} = N\gamma$. Thus we find
\begin{equation}
\eta_{j}= \gamma = \frac{\bar{\eta}\,\eta_{0}^{\,2}}{\eta_{0}^{\,2}+N\bar{\sigma}^2\left[1+(N-1)\alpha\right]}. 
\end{equation}
In addition, the optimal total density $\eta_{\textrm{opt}}$ is given by
\begin{equation}\label{EQN:EtaOpt}
\eta_{\textrm{opt}}= N\gamma = N\frac{\bar{\eta}\eta_{0}^{\,2}}{\eta_{0}^{\,2}+N\bar{\sigma}^2\left[1+(N-1)\alpha\right]}. 
\end{equation}

There is another way one can derive this last result [Eq.~(\ref{EQN:EtaOpt})], which is helpful in understanding the nature of the optimization being carried out, and so we outline the results of this alternate derivation here. Namely, when $P(\vec{\eta}\,|\mathcal{T})$ is Gaussian with mean vector $(\eta_{1}^{\star}, \eta_{2}^{\star},\dots,\eta_{N}^{\star})$ and covariance matrix $\mathcal{C}$ [as in Eq.~(\ref{EQN:CorrGauss})], then the probability distribution over the sum $\sum_{i=1}^{N}\eta_{i}\equiv \eta$, denoted by $R(\eta)$, is also Gaussian with mean $\sum_{i=1}^{N}\eta_{i}^{\star}$ and variance $\sum_{i,j=1}^{N}\mathcal{C}_{ij}$ (see for example~\cite[chapter~II, Sec. 13]{shiryaev_96}). Under the simplifying assumptions introduced earlier, namely, if we set $\eta_{i}^{\star}=\bar{\eta}$ for all $i$, and $\mathcal{C}\to\tilde{\mathcal{C}}$ as in Eq.~(\ref{EQN:CovarianceMatrix}), we have: $\sum_{i=1}^{N}\eta_{i}^{\star}=N\bar{\eta}$ and $\sum_{i,j=1}^{N}\mathcal{C}_{ij} = N\bar{\sigma}^2\left[1+(N-1)\alpha\right]$. Thus
\begin{equation}
R(\eta)\propto\exp\left\{-\frac{1}{2N\bar{\sigma}^2\left[1+(N-1)\alpha\right]}(\eta-N\bar{\eta})^2\right\}.
\end{equation}
The resulting probability distribution over the sum of the densities $\eta$, denoted by $P_{\textrm{tot}}(\eta|\mathcal{T}, W)$, can be shown to be:
\begin{widetext}
\begin{eqnarray}
P_{\textrm{tot}}(\eta|\mathcal{T}, W) &\propto& R(\eta)W(\eta)\nonumber \\
&=& \exp\left\{-\frac{1}{2N\bar{\sigma}^2\left[1+(N-1)\alpha\right]}(\eta-N\bar{\eta})^2\right\}\exp\left\{-\frac{1}{2\eta_{0}^{\,2}}\eta^2\right\}\nonumber\\
&=&\exp\left\{-\frac{1}{2\Sigma^{2}}(\eta-\Phi)^2\right\}, \label{EQN:PtotEtot}
\end{eqnarray}
\end{widetext}
where
\begin{eqnarray}
\Sigma^2 &=& \frac{\eta_{0}^{\,2}N\bar{\sigma}^2\left[1+(N-1)\alpha\right]}{\eta_{0}^{\,2}+N\bar{\sigma}^2\left[1+(N-1)\alpha\right]},\label{EQN:PtotEtotSIGMA}
 \\
\Phi &=& N\frac{\bar{\eta}\eta_{0}^{\,2}}{\eta_{0}^{\,2}+N\bar{\sigma}^2\left[1+(N-1)\alpha\right]}.\label{EQN:PtotEtotCENTER}
\end{eqnarray}
Thus $P_{\textrm{tot}}(\eta|\mathcal{T}, W)$ is also Gaussian with its maximum occurring at $\Phi$; in agreement with the optimal value $\eta_{\textrm{opt}}$  displayed in Eq.~(\ref{EQN:EtaOpt}). 

\subsection{Prediction and fine tuning}\label{SEC:PredFineTuning}

How then, in light of the above discussion, do we propose to extract a prediction from $P_{\textrm{tot}}(\eta\,|\mathcal{T}, W)$ while allowing for variations in assumptions regarding typicality? We know that the probability distribution $P_{\textrm{tot}}(\eta\,|\mathcal{T}, W)$ is Gaussian [see Eqs.~(\ref{EQN:PtotEtot}),~(\ref{EQN:PtotEtotSIGMA}), and~(\ref{EQN:PtotEtotCENTER})], and so it has a single maximum; moving sufficiently far away from this maximum takes us into regions of atypicality. Hence, introducing a factor $F>0$ that measures deviations from the maximum, we propose that the framework specified by the theory $\mathcal{T}$, together with the anthropic conditionalization factor $W(\eta)$, and the typicality assumption characterized by $F$, is
\begin{equation}\label{EQN:Predictive}
\textrm{\emph{predictive} if}\;\;\frac{1}{F}\eta_{\textrm{opt}} = \eta_{\textrm{obs}}, 
\end{equation}
where $F=1$ corresponds to the assumption of `maximum' typicality, and deviations from $F=1$ correspond to some degree of atypicality.

In addition, following~\citet{aguirre+tegmark_05}, we do not want this prediction to be too finely tuned, in the sense that increasing the value of the prediction (i.e., $\frac{1}{F}\eta_{\textrm{opt}}$) should not take us too far into the tail of the anthropic conditionalization factor $W(\eta)$ [given by Eq.~(\ref{EQN:AnthropicFactor})]. In this way, we will assume that the value of the prediction for the total observed dark matter density is
\begin{equation}\label{EQN:NFT}
\textrm{\emph{not finely tuned} if}\;\;\frac{1}{F}\eta_{\textrm{opt}} \leq 2\eta_{0},
\end{equation}
namely, within two standard deviations, $2\eta_{0}$, of the mean of the Gaussian conditionalization factor $W(\eta)$. The precise tolerance here is less important than the general conclusions we will develop below. 

We will show (as indeed mentioned by~\citet{aguirre+tegmark_05}) that the criteria expressed in Eqs.~(\ref{EQN:Predictive}) and~(\ref{EQN:NFT}) can be used to predict the total number of equally contributing components to the total dark matter density, and it is within the context of this type of prediction that we will analyze the effects of atypicality. To gain intuition about how these two criteria operate, we begin by analyzing the case of independent species of dark matter.

\subsection{Independent species and typicality assumptions}

In the case of independent species of dark matter, namely, when $\alpha=0$, a quick calculation reveals that the prediction that $\frac{1}{F}\eta_{\textrm{opt}}=\eta_{\textrm{obs}}$ [Eq.~(\ref{EQN:Predictive})], in combination with Eq.~(\ref{EQN:EtaOpt}), can be translated into a prediction for $N$, with
\begin{equation}\label{EQN:NGaussPredictive}
N = \frac{F\eta_{\textrm{obs}}\eta_{0}^{\,2}}{\eta_{0}^{\,2}\bar{\eta}-F\eta_{\textrm{obs}}\bar{\sigma}^{2}}. 
\end{equation}
The demand that the original prediction is not finely tuned, namely, that $\frac{1}{F}\eta_{\textrm{opt}}\leq 2\eta_{0}$ [Eq.~(\ref{EQN:NFT})], bounds $N$ such that
\begin{equation}\label{EQN:NGaussAccept}
N \leq \frac{2F\eta_{0}^{\,2}}{\eta_{0}\bar{\eta}-2F\bar{\sigma}^{2}},  
\end{equation}
for $\eta_{0}\bar{\eta} > 2F\bar{\sigma}^{2}$; otherwise no such upper bound exists. Equations~(\ref{EQN:NGaussPredictive}) and~(\ref{EQN:NGaussAccept}) imply that if $0<F<1$, both the value of $N$ that is predictive and the upper bound on $N$ such that the prediction is not finely tuned, decrease relative to the case of typicality (i.e., relative to $F=1$). When $F>1$, these values increase relative to the case of typicality. In this way, atypicality can change the nature of the prediction. 

\subsection{Correlated species and typicality assumptions}

For nonzero $\alpha$, again using Eq.~(\ref{EQN:EtaOpt}), we find that the framework we are examining is predictive, i.e., satisfies Eq.~(\ref{EQN:Predictive}), when 
\begin{equation}\label{EQN:DGaussPredictive}
\alpha\eta_{\textrm{obs}}\bar{\sigma}^{2}N^2 + \left[(1-\alpha)\eta_{\textrm{obs}}\bar{\sigma}^{2}-\frac{1}{F}\bar{\eta}\eta_{0}^{\,2}\right]N+\eta_{\textrm{obs}}\eta_{0}^{\,2} = 0, 
\end{equation}
and this prediction is not finely tuned, i.e., satisfies Eq.~(\ref{EQN:NFT}), when
\begin{equation}\label{EQN:DGaussAcceptable}
-2\alpha\bar{\sigma}^{2}N^2 + \left[\frac{1}{F}\bar{\eta}\eta_{0}-2\bar{\sigma}^{2}(1-\alpha)\right]N-2\eta_{0}^{\,2}\leq 0.
\end{equation}
We note that now, depending on the balance of the parameters in the problem, it is possible that under the inclusion of nonzero correlations, there exist two distinct solutions to the prediction for the total number of species that contribute equally to the total dark matter density. We focus first on this effect in more detail, before studying the effects of atypicality on the resulting predictions. In particular, we will be interested in two different questions: (1) under the assumption of typicality, how does a nonzero $\alpha$ change predictions for $N$ (relative to the $\alpha=0$ case, and assuming these predictions are not finely tuned), and (2) for a fixed nonzero $\alpha$, how does atypicality change the prediction for $N$ (again assuming these predictions are not finely tuned)?

\subsubsection{The effect of correlations, $\alpha\neq 0$, on $N$, under typicality}\label{SEC:CorrEffectonN}

In addressing the first question, we are interested in comparing, for $F=1$, Eq.~(\ref{EQN:NGaussPredictive}) and the solution(s) to Eq.~(\ref{EQN:DGaussPredictive}). To make the comparison more tractable, we make two simplifying assumptions: namely, (i) that the original probability distribution $P(\vec{\eta}|\mathcal{T})$ has significant probability near the observed value; more precisely, we will assume
\begin{equation}\label{EQN:etaEQetaobs}
\bar{\eta}\equiv\eta_{\textrm{obs}},
\end{equation}
and (ii) that the variance of the conditionalization factor $W(\eta)$ is related in a simple way to the variances of the dark matter components, 
\begin{equation}\label{EQN:Xrelation}
\eta_{0}^{\,2} = X\bar{\sigma}^{2},
\end{equation}
for some positive $X$ whose range will be specified shortly. 

Under these assumptions then, the prediction for the number of uncorrelated species, as derived from Eq.~(\ref{EQN:NGaussPredictive}), and which we will now refer to as $N_{0}$, is
\begin{equation}\label{EQN:NGaussPredictiveRED}
N_{0} = \frac{X}{X-1}. 
\end{equation}
In order that $N_{0}$ is a physically realizable prediction (and so that we do not, at this stage, discount the framework that gave rise to this prediction), we choose $X>1$, so that $N_{0}>1$.

Similarly, these assumptions imply that Eq.~(\ref{EQN:DGaussPredictive}) reduces to 
\begin{equation}\label{EQN:PredictiveParabolaRED}
\alpha N_{\alpha}^2+(1-\alpha-X)N_{\alpha}+X=0,
\end{equation}
where we now denote $N$ by $N_{\alpha}$; the solution of which is
\begin{equation}\label{EQN:DGaussPredictiveRED}
N_{\alpha} = \frac{X+\alpha-1\pm\sqrt{(1-\alpha-X)^2-4\alpha X}}{2\alpha}. 
\end{equation}

\begin{figure*}
\includegraphics[width=1\linewidth]{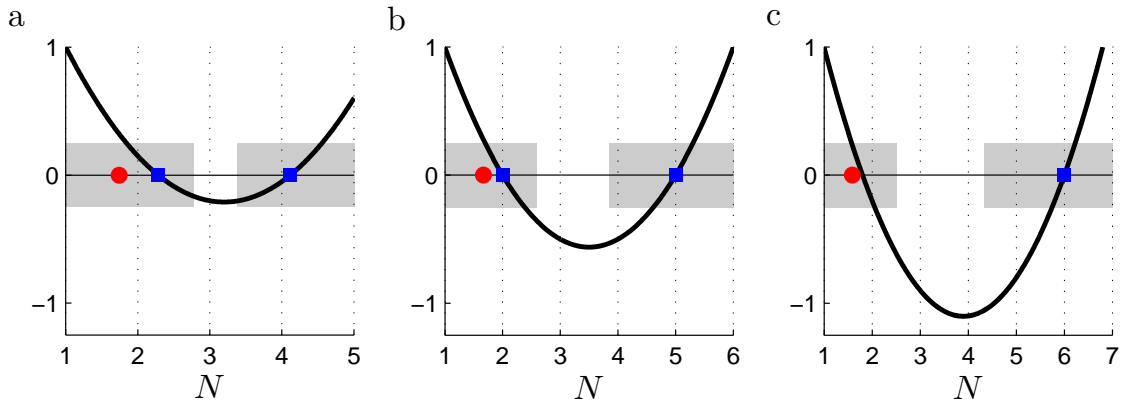}
\hfill
\caption{Change in the prediction for $N$ in the presence of correlations under the assumption of typicality. (a,b,c) exhibit solutions for the choices $X=2.35, 2.5,\textrm{and}\,2.7$ respectively. The red circles correspond to $N_{0}$, the prediction for independent species of dark matter [see Eq.~(\ref{EQN:NGaussPredictiveRED})---each of these solutions is not finely tuned under the choice of parameters described herein]. The parabolas correspond to the left-hand side of Eq.~(\ref{EQN:PredictiveParabolaRED}), whose $x$-axis-intercepts are the predictions for $N_{\alpha}$, shown in blue squares, where $\alpha=0.25$ in each case [a prediction of $N_{\alpha}<2$, as for the smaller of the two predictions in (c), is discounted as `unphysical']. The gray segments overlapping the $x$-axes correspond to the range of solutions that are not finely tuned for the correlated case (namely, the region of the $x$-axis where the constraint given by Eq.~(\ref{EQN:DGaussAcceptable}) is satisfied---recall that $F=1$ under the assumption of typicality, and we have set $\bar{\eta}=\eta_{\textrm{obs}}$, and $\eta_{0}^{\,2} = X\bar{\sigma}^{2}$). We have set $\eta_{\textrm{obs}}=5$ in accord with the experimentally observed value, and for the sake of illustration, we have set $\bar{\sigma}^2=2.8$. We note that in (a,b), there exist two, distinct, physically acceptable predictions for $N_{\alpha}$, the greater of which is also significantly different from the case where there are no correlations.}
\label{FIG:AlphaEffect}
\end{figure*}

Let us note a couple of cases of interest here. Firstly, if $(1-\alpha-X)^2=4\alpha X$, then there is just a single solution $N_{\alpha} = (X+\alpha-1)/{2\alpha}$. Setting $\alpha=0.25$ for example, gives $X=2.25$ and so $N_{\alpha}=3$ whereas $N_{0} \approx 2$ (one can show that both of these solutions are not necessarily finely tuned). In the case that $(1-\alpha-X)^2>4\alpha X$, $N_{0}$ will exhibit just a single solution, whereas $N_{\alpha}$ may exhibit two (physical) solutions. Figure~\ref{FIG:AlphaEffect} displays some illustrative examples of what these solutions look like. There we exhibit solutions $N_{0}$ (red circles) and $N_{\alpha}$ (blue squares) for $\alpha=0.25$, under the assumption that $X=2.35, 2.5,\textrm{or}\,2.7$ (corresponding to Fig.~\ref{FIG:AlphaEffect}a,~\ref{FIG:AlphaEffect}b, or~\ref{FIG:AlphaEffect}c respectively). We see that in Figs.~\ref{FIG:AlphaEffect}a and~\ref{FIG:AlphaEffect}b, the introduction of correlations leads to two distinct solutions for $N_{\alpha}$, the greater of which is significantly different from $N_{0}$. In the case of Fig.~\ref{FIG:AlphaEffect}c, the smaller of the two solutions for $N_{\alpha}$  is discounted as unphysical (as only those solutions in the correlated case make sense where $N_{\alpha}\geq 2$).
\subsubsection{How does atypicality change the prediction when $\alpha\neq 0$?}

The second question we are interested in is the nature of the change in the prediction as a result of atypicality in the correlated Gaussian case. To investigate this in a simple setting, we again invoke the assumptions of section~\ref{SEC:CorrEffectonN} as expressed in Eqs.~(\ref{EQN:etaEQetaobs}) and~(\ref{EQN:Xrelation}). The equation expressing predictivity, Eq.~(\ref{EQN:DGaussPredictive}), reduces to
\begin{equation}
\alpha N_{\alpha}^2+(1-\alpha-\frac{1}{F}X)N_{\alpha}+X=0,
\end{equation}
and the bound on $N_{\alpha}$ such that the prediction is not finely tuned, namely Eq.~(\ref{EQN:DGaussAcceptable}), reduces to
\begin{equation}
-2\alpha\bar{\sigma}N_{\alpha}^2 + \left[\frac{1}{F}\eta_{\textrm{obs}}\sqrt{X}-2\bar{\sigma}(1-\alpha)\right]N_{\alpha}-2X\bar{\sigma}\leq 0.
\end{equation}

\begin{figure*}
\includegraphics[width=1\linewidth]{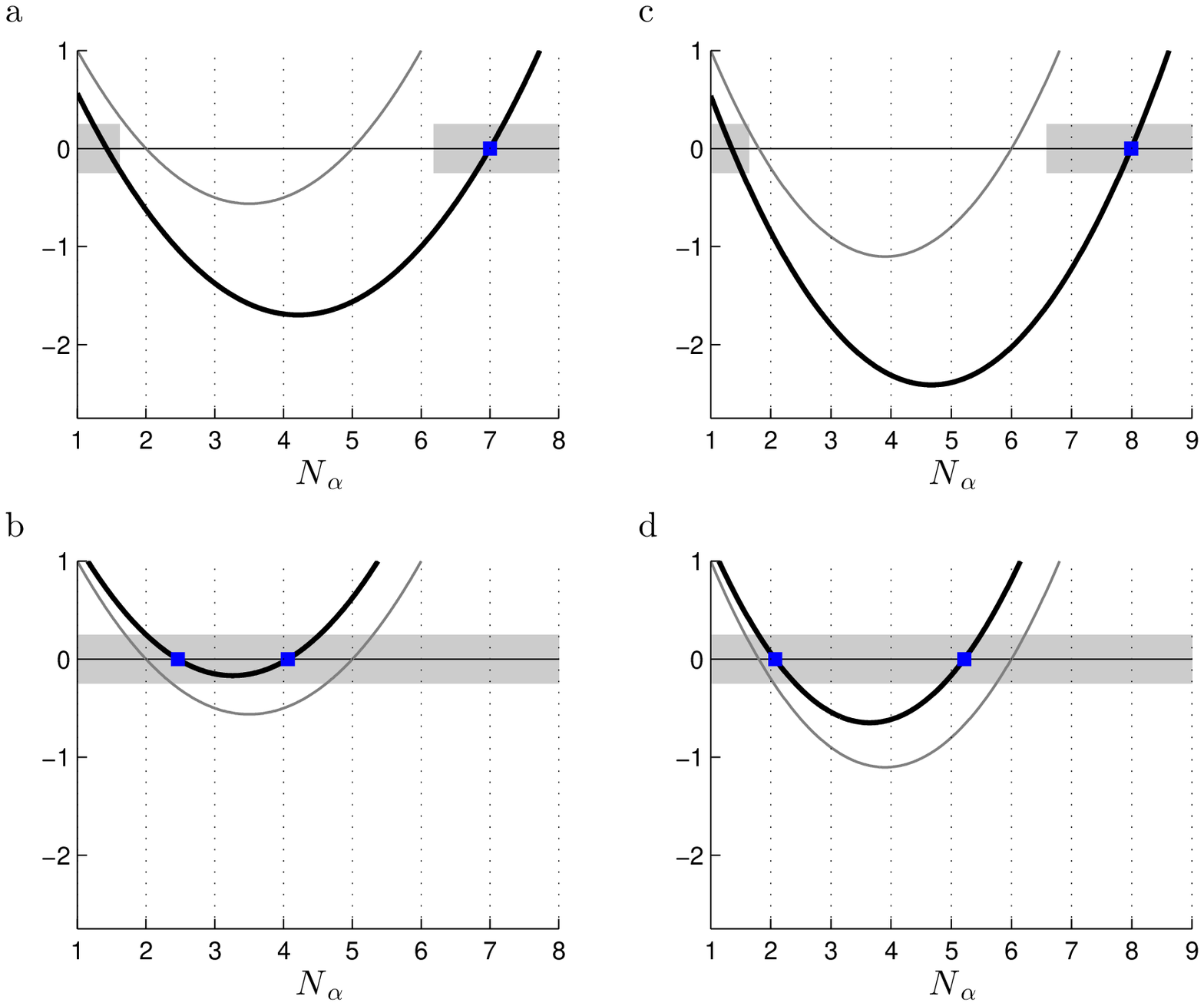}
\hfill
\caption{The effects of atypicality on the prediction of $N_{\alpha}$, where $\alpha=0.25$. (a) $X=2.5$, $F = 0.875$; (b) $X=2.5$, $F = 1.05$; (c) $X=2.7$, $F = 0.875$; (d) $X=2.7$, $F = 1.05$. In each case, we have also set $\eta_{\textrm{obs}}=5$ and $\bar{\sigma}^2=2.8$. For each panel, the intersection of the gray parabola with the $x$-axis corresponds to the prediction under typicality, whereas the intersection of the black parabola with the $x$-axis, marked by blue squares, corresponds to the prediction under the assumption of atypicality (recall that we accept only those solutions for which $N_{\alpha}\geq 2$). The gray segment overlapping the $x$-axis corresponds to the range of solutions for the atypical scenario that are not finely tuned. We see in each case that predictions for $N_{\alpha}$ can shift significantly.}
\label{FIG:TypEffect}
\end{figure*}

For illustrative values of the parameters, the effects of these equations on the prediction of the total number of species of dark matter contributing equally to the total dark matter density are explored in Fig.~\ref{FIG:TypEffect}. We see there that the prediction under the assumption of atypicality [as determined by the appropriate $x$-axis-intercept(s) of the black parabola in each panel; recall that only those solutions where $N_{\alpha} \geq 2$ are considered physical] changes significantly from the prediction under the assumption of typicality [corresponding to the appropriate $x$-axis-intercept(s) of the gray parabola in each panel]. 

\section{Different frameworks, same prediction}\label{SEC:Overlaps}

So let us recapitulate what we have found thus far, in order to better understand the nature of the overlaps that exist between different frameworks as regards their predictions for the total number of species of dark matter that we should expect to observe.

In the case of bottom-up conditionalization (Sec.~\ref{SEC:BU}), where we assumed the underlying probability distribution $P(\vec{\eta}|\mathcal{T})$ was unimodal and could, in principle, take nonzero values within some $N$-dimensional cube in parameter space, the expected number of dominant dark matter components, under the assumption of typicality, was shown to be 1. In the terminology of Sec.~\ref{SEC:BU}, $\langle j\rangle\sim 1$ for $M\gg N$. However, atypicality can change this prediction to a range of other possibilities, including equal contributions from all $N$ components. 

In the case of top-down conditionalization (Sec.~\ref{SEC:TD}), we found the opposite prediction, that for a correlated Gaussian distribution $P(\vec{\eta}|\mathcal{T})$, all $N$ components, under typicality, contribute significantly to the total dark matter density [Eq.~(\ref{EQN:EqualContribution})]. This result holds for probabilistically independent ($\alpha = 0$) dark matter species as well; and it generalizes the previous result of~\citet[Sec.~3.3]{aguirre+tegmark_05} for a particular $P(\vec{\eta}|\mathcal{T}$). However, there exist regions of parameter space such that assumptions of atypicality lead to the prediction of just a single dominant dark matter component [generalizing the argument in~\citet[Sec.~3.2]{azhar_14}, again, for a particular $P(\vec{\eta}|\mathcal{T})$]. In this case also, we found that non-Gaussianities can overturn the equality of contribution, such that typicality corresponds to a single dominant species (for the $N=2$ case studied there), and atypicality corresponds to equal contributions, or indeed to a single species dominating to a greater degree than in the case of typicality (as in Fig.~\ref{FIG:NonGaussOptimization}c and~\ref{FIG:NonGaussOptimization}f). 

Finally, for the anthropic case (Sec.~\ref{SEC:AR}), we explored the assumption of atypicality in a different way to the first two approaches: namely, by tracking its impact on the total number $N$ of equally contributing components (indeed, in Sec.~\ref{SEC:AR}, $N$ could vary, unlike in the bottom-up and top-down cases where it was fixed by assumption at the outset). Now, under typicality, correlations in the underlying probability distribution can change the prediction for $N$ relative to the independent case (see Fig.~\ref{FIG:AlphaEffect}), and it is possible for two physically acceptable predictions to exist, which again, can change quantitatively under the assumption of atypicality (as in Fig.~\ref{FIG:TypEffect}).

It is evident from the above discussion that the types of prediction discussed here do not cleanly discriminate between frameworks consisting of theory, conditionalization scheme and typicality assumption. For example, consider first the prediction that dark matter consists of a single dominant component. This could be derived from each conditionalization scheme studied above in (at least) the following ways: 
\begin{itemize}
\item[---]\emph{bottom-up}: for an $N$-dimensional unimodal distribution under typicality (Sec.~\ref{SEC:BU}); 
\item[---]\emph{top-down}: for uncorrelated or correlated $N$-dimensional Gaussians under  atypicality [Eq.~(\ref{EQN:AtypRatio}), and items (i) and (ii) at the end of Sec.~\ref{SEC:TDGaussAtyp}], or the 2-dimensional non-Gaussian distribution of Eq.~(\ref{EQN:NonGaussDist}) under typicality (Fig.~\ref{FIG:NonGaussOptimization}c and \ref{FIG:NonGaussOptimization}f); 
\item[---]\emph{anthropic}: for uncorrelated Gaussians assuming typicality, Eqs.~(\ref{EQN:etaEQetaobs}),~(\ref{EQN:Xrelation}), and $X\gg 1$ [so that Eq.~(\ref{EQN:NGaussPredictiveRED}) implies $N_{0}\sim 1$].
\end{itemize}

The prediction of multiple species of dark matter does not fare any better in terms of its ability to discriminate between frameworks. Consider the prediction of two equally dominant species of dark matter. This could also be derived from each conditionalization scheme in (at least) the following ways:
\begin{itemize}
\item[---]\emph{bottom-up}: for an $N$-dimensional unimodal distribution under an appropriate assumption of atypicality (as discussed at the end of Sec.~\ref{SEC:BU}); 
\item[---]\emph{top-down}: for correlated 2-dimensional Gaussians under typicality (see Fig.~\ref{FIG:NonGaussOptimization}a and~\ref{FIG:NonGaussOptimization}d), or the 2-dimensional non-Gaussian distribution of Eq.~(\ref{EQN:NonGaussDist}) under atypicality (see Fig.~\ref{FIG:NonGaussOptimization}c and \ref{FIG:NonGaussOptimization}f); 
\item[---]\emph{anthropic}: for correlated Gaussians under typicality (as in the smaller of the two distinct predictions of Fig.~\ref{FIG:AlphaEffect}a and~\ref{FIG:AlphaEffect}b) or from the smaller of two distinct predictions under atypicality (as in Fig.~\ref{FIG:TypEffect}b and~\ref{FIG:TypEffect}d). 
\end{itemize}

In this sense, distinct frameworks can overlap as regards their predictions. This possibility raises difficulties for how we can confirm frameworks in cosmological models of the multiverse---as we discuss in the next section. 

\section{Discussion}\label{SEC:Discussion}

For theories that describe a multiverse, the confirmation of these theories must---short of direct experimental evidence---rest on tests such as those explored in this paper. Any such theory will probably describe an overwhelming number of domains that look nothing like ours, in which case theory alone will not be enough to extract meaningful predictions. Indeed, conditionalization will be needed, in which we restrict attention to domains in such a way as to sharpen the comparison between what the theory predicts and what we observe. 

Any of the conditionalization schemes outlined by~\citet{aguirre+tegmark_05} and studied herein, or indeed more sophisticated versions of these, are plausible candidates; but there is an inherent arbitrariness in the choice. And as explored in this paper, the situation is further complicated by assumptions regarding typicality. 

The argument that we need to question typicality in multiverse settings has been made elsewhere~\cite{hartle+srednicki_07, srednicki+hartle_10, azhar_14, azhar_15, hartle+hertog_15}; but it is helpful to rehearse central features of that argument to more clearly grasp the motivations that underlie this paper. The main point is that, although our observational situation might be unlikely according to theories of the multiverse, they may well posit \emph{multiple} domains in which our observational situation exists---as described by a conjunction of a theory and some appropriate conditionalization scheme. And if observables (such as the outcomes of future experiments) can take different values in these domains, the appropriate test of the conjunction will be a comparison of what we observe with what the conjunction predicts for \emph{our} observations (a first-person prediction, in the terminology of~\citet{srednicki+hartle_10}). We cannot know, of course, which of these domains we are in and so to extract an appropriate prediction, we need to make an assumption about our typicality with respect to these domains. Under such circumstances, the assumption that we are typical is certainly not guaranteed a priori; and so it makes sense to then allow for a variety of assumptions regarding typicality in order to identify the most predictive framework~\cite{srednicki+hartle_10, azhar_15}.

In this paper we have taken seriously the conclusions of the last paragraph, and have applied them to the comparison of the total number of observed species of dark matter with predictions generated from some theory of the multiverse. A central feature has been that what such a comparison tests is an entire \emph{framework}, namely, a conjunction of theory, conditionalization scheme, and typicality assumption. Hence if the prediction of such a conjunction does not match our observations, we must disfavor the entire conjunction; and thus we have license to change any of its conjuncts, and to then reassess the predictive power of the resulting  framework. What we find under this scenario, as argued in this paper, is a complex set of interconnected relationships between frameworks and predictions. Indeed, as drawn out in Sec.~\ref{SEC:Overlaps}, the same prediction can arise from distinctly different frameworks.

It would be interesting to see how widespread these `overlaps' are for more realistic cosmological scenarios. If they are also robust to the choice of the physical observables we aim to predict the values of, and we believe that truly distinct frameworks indeed give rise to the same prediction, then we are forced to conclude that the prediction cannot confirm any single framework taken on its own. Of course, one has recourse to more intricate confirmation schemes, such as those invoked in Bayesian analyses---which would introduce priors over frameworks to help in their demarcation (see~\cite{hartle+srednicki_07, srednicki+hartle_10} for example). But in the context where we focus solely on likelihoods (as we have implicitly done in this paper), robust overlaps between frameworks present an acute challenge for the utility of such tests of the multiverse.

\begin{acknowledgments}
I am very grateful to Jeremy Butterfield for discussions and comments on an earlier version of this paper, and to Jim Hartle for posing a question over email at an early stage, which helped to guide my thoughts along the general lines expressed in this paper. I am supported  by the Wittgenstein Studentship in Philosophy at Trinity College, Cambridge.
\end{acknowledgments}

\bibliography{azhar_15b_BIB.bib}

\end{document}